\documentclass[showpacs,prd,a4paper,nofootinbib]{revtex4}

\usepackage{braket}
\usepackage{amsfonts}
\usepackage{amsmath}
\usepackage{bbm}
\usepackage{bm}
\usepackage{graphicx}
\usepackage{color}

\newcommand{\Dslash}{D\!\!\!\! /}
\newcommand{\ud}{\mathrm{d}}
\newcommand{\uD}{\mathcal{D}}
\newcommand{\Tr}{\mathrm{Tr}}
\newcommand{\condsigmap}{\braket{\sigma^\prime}}
\newcommand{\condazpmp}{\braket{a_0^{\pm \prime}}}
\newcommand{\condazp}{\braket{\bm{a}_0^\prime}}
\newcommand{\condazzp}{\braket{a_0^{0 \prime}}}
\newcommand{\condetap}{\braket{\eta'}}
\newcommand{\condpipmp}{\braket{\pi^{\pm \prime}}}
\newcommand{\condpip}{\braket{\bm{\pi}^\prime}}
\newcommand{\condpizp}{\braket{\pi^{0 \prime}}}
\newcommand{\half}{\frac{1}{2}}

\begin{document}

\title{Spontaneous CP-violation in the strong interaction at $\theta=\pi$}

\author{Dani\"el Boer}
\author{Jorn K. Boomsma}
\affiliation{Department of Physics and Astronomy, Vrije Universiteit 
Amsterdam\\
De Boelelaan 1081, NL-1081 HV Amsterdam, the Netherlands}

\date{\today}

\begin{abstract}
  Spontaneous CP-violation in the strong interaction is analyzed at $\theta =
  \pi$ within the framework of the two-flavor NJL model. It is found that the
  occurrence of spontaneous CP-violation at $\theta = \pi$ depends on the
  strength of the 't Hooft determinant interaction, which describes the effect
  of instanton interactions.  The dependence of the phase structure, and in
  particular of the CP-violating phase, on the quark masses, temperature,
  baryon and isospin chemical potential is examined in detail. When available
  a comparison to earlier results from chiral perturbation theory is made.
  From our results we conclude that spontaneous CP-violation in the strong
  interaction is an inherently low-energy phenomenon. In all cases we find
  agreement with the Vafa-Witten theorem, also at nonzero density and
  temperature. Meson masses and mixing in the CP-violating phase display some
  unusual features as a function of instanton interaction strength. A
  modification of the condition for charged pion condensation at nonzero
  isospin chemical potential and a novel phase of charged $a_0$ mesons are
  discussed.
\end{abstract}



\maketitle
\section{Introduction}
The possibility of CP-violation in the strong interaction has been
studied extensively. The QCD Lagrangian naturally incorporates a
$\theta$-term $\mathcal{L}_\theta = \frac{\theta g^2}{32 \pi^2} F \tilde
F$, which can lead to CP-violation due to instanton contributions.
Only for $\theta = 0~(\rm{mod}\ \pi)$ the Lagrangian is CP-conserving.
In nature the value of $\theta$ is extremely close to zero, as has
been concluded from pseudoscalar mass ratios and the neutron electric
dipole moment~\cite{Baluni:1978rf,
  Crewther:1979pi,Kawarabayashi:1980uh}. This suggests that
$\theta=0$, but the lack of a satisfactory explanation of why this
should be the case is commonly referred to as the strong CP problem.

At $\theta=0$ no explicit CP-violation is present in the Lagrangian, but in
addition, the well-known Vafa-Witten theorem states that spontaneous parity
violation in QCD at $\theta=0$ does not arise~\cite{Vafa:1984xg}. This rules out
the possibility that $\braket{F \tilde F} \neq 0$ at $\theta=0$ and implies
that the QCD ground state for nonzero $\theta$ must have higher energy than at
$\theta=0$.  The situation is different at $\theta = \pi$, when the Lagrangian
is also explicitly CP-conserving. In this case spontaneous CP-violation could
arise, as was first pointed out by Dashen~\cite{Dashen:1970et}.  There are then
two degenerate CP-violating vacua, which differ by a CP transformation from
each other.

\begin{figure}[htb]
  \includegraphics[width=0.4\textwidth]{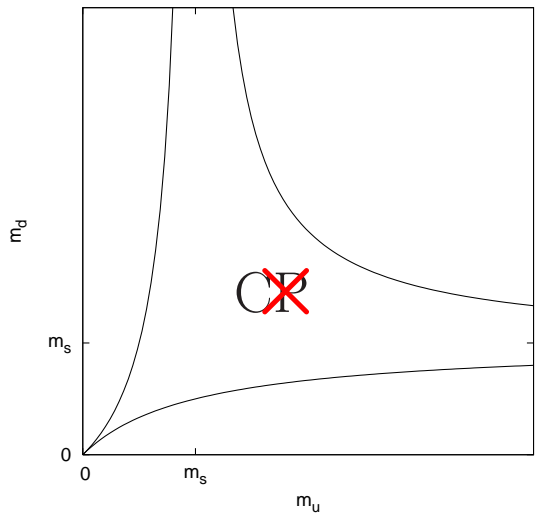}
  \caption{The $(m_u,m_d)$ phase diagram of three-flavor chiral
  perturbation theory, with $\theta = \pi$ and $a/N \to \infty$, cf.\ 
  Ref.\ \cite{Creutz:2003xu}.}
  \label{ChPT_NF3_mu_md}
\end{figure}

This possibility of spontaneous CP-violation is one of the reasons why people
have studied the $\theta$-dependence of the strong interactions. However, to
study this in full QCD is very difficult due to the nonperturbative nature of
the $\theta$-term. Even in lattice QCD, studies are limited to small $\theta$,
because of the problem of how to deal with complex phases. Therefore, the
  $\theta$-dependence of the strong interaction and Dashen's phenomenon have
  been studied extensively using low energy effective theories, such as chiral
  perturbation
  theory~\cite{Witten:1980sp,DiVecchia:1980ve,Smilga:1998dh,Tytgat:1999yx,Akemann:2001ir,Creutz:2003xu,Metlitski:2005db,Metlitski:2005di},
  or by using specific models, such as the NJL model~\cite{fujihara:2005wk}.
In a quark model, like the NJL model, the effects of instantons and the
$\theta$-term are incorporated via an effective interaction, the 't Hooft
determinant interaction~\cite{tHooft:1976fv,tHooft:1986nc}. In chiral perturbation
theory these effects can be included in a similar way via a log determinant
interaction~\cite{Witten:1980sp,DiVecchia:1980ve}.  Here we shall denote the
strength of the latter interaction by $a$.  Whether or not the theory exhibits
spontaneous CP-violation at $\theta = \pi$ depends on $a$ and on the values of
the quark masses.  Two limiting cases were discussed in the literature.
Ref.~\cite{Witten:1980sp}, which considers lowest-order chiral perturbation
theory, states that when $a/N$ is nonzero but much smaller than the quark
masses, the theory always exhibits spontaneous CP-violation at $\theta = \pi$,
independent of the values of the quark masses and number of flavors $N$. The
opposite case \cite{Witten:1980sp,DiVecchia:1980ve,Tytgat:1999yx}, 
i.e.\ when the masses of the quarks are much smaller
than $a/N$, leads to different results. In this case it does depend on the 
values of the quark masses. In the two-flavor case for $a/N \to \infty$ 
(which means no $\eta$ meson is included), spontaneous CP-violation only
occurs for degenerate quark masses (in that case actually for all finite 
values of $a/N$). For finite $a/N \gg m_u,m_d$ spontaneous 
CP-violation also occurs for nondegenerate quark masses in a finite interval
of $m_d/m_u$ around 1, as was shown in Ref.\ \cite{Tytgat:1999yx}.
In the three-flavor case, a region exists
in the $(m_u,m_d)$-plane where the theory spontaneously violates CP
invariance~\cite{Creutz:2003xu}, as shown in Fig.~\ref{ChPT_NF3_mu_md}. The
asymptotes depend on the value of the strange quark mass.

In first-order chiral perturbation theory there are only a few parameters,
namely the quark masses, the pion decay constant and the strength of the
determinant interaction. It is therefore interesting to study CP-violation in
a somewhat richer situation, such as chiral perturbation theory beyond leading
order, which has been studied in Ref.\ \cite{Smilga:1998dh} for $a/N \to
\infty$. 
Here we will make a comprehensive study of the $\theta$-dependence and
especially spontaneous CP-violation at $\theta = \pi$ within the framework of
the two-flavor NJL model in the mean-field approximation.  We will study the
dependence on the effective instanton interaction strength (denoted by $c$ in
this case), not only in the two limiting cases, but for all possible values.
We find that there is a critical value of the interaction strength above
  which spontaneous CP violations occurs and which depends linearly on the
quark masses, as expected from axial anomaly considerations.

As will be discussed, the two-flavor NJL model allows for spontaneous
CP-violation also for nondegenerate quark masses. We
find a region in the $(m_u,m_d)$-plane very similar to the three-flavor
lowest-order chiral perturbation theory result shown in
Fig.~\ref{ChPT_NF3_mu_md}. However, for the two-flavor NJL model the
asymptotes are determined by the strength of the instanton induced
interaction.

We will also study the influence of nonzero temperature and baryon and
isospin chemical potential. It has been suggested that in those cases
the Vafa-Witten theorem may no longer apply (see for instance Ref.\
\cite{Cohen:2001hf} for some explicit arguments, but also Ref.\
\cite{Einhorn:2002rm} for counterarguments). But even if it does
apply, spontaneous CP-violation at finite temperature or baryon
chemical potential (through meta-stable states) has been considered in
the literature
\cite{Lee:1973iz,Morley:1983wr,Kharzeev:1998kz,Buckley:1999mv} and
possible experimental signatures in heavy ion collisions have been put
forward
\cite{Kharzeev:1999cz,Voloshin:2004vk,Kharzeev:2007tn,Kharzeev:2007jp}.
This would also be relevant in the early universe, when possibly $\theta$
was nonzero and later relaxed to zero, for example via a Peccei-Quinn-like
mechanism~\cite{Peccei:1977hh,Peccei:1977ur,Wilczek:1977pj}. 
We therefore wish to check the Vafa-Witten theorem and the possible
presence of CP-violating local minima in the NJL model at finite
temperature and density.

Spontaneous CP-violation at $\theta = \pi$ within the two-flavor NJL
model including temperature dependence has been considered before in
Ref.~\cite{fujihara:2005wk}, but only for a very limited range of quark
masses: $|m_u \pm m_d| < 6$ MeV and without chemical potentials.
 
In Ref.~\cite{Metlitski:2005di} the phase diagram as a function of $\theta$
and isospin chemical potential has been investigated within
first-order chiral perturbation theory for two flavors and effectively
$a/N \to \infty$ (due to the absence of the $\eta$ meson). We will
compare this to our results at nonzero isospin chemical potential,
where a modification of the pattern of charged pion condensation is
observed at $\theta=\pi$.

This article is organized as follows. First the NJL model is briefly
introduced to set the notation, then we discuss the effect of chiral
transformations on the theory, which is relevant for the calculation
of the effective potential and for a comparison to earlier results
from the literature. We continue with a discussion of the
$\theta$-dependence of the ground-state, including temperature effects
and nonzero baryon and isospin chemical potential. Also we discuss the 
$c$-dependence of the meson masses and mixing in the CP-violating phase. 
We end with conclusions and a further discussion of the results.

\section{The NJL model}
The Nambu-Jona-Lasinio (NJL) model, introduced in
Refs.~\cite{Nambu:1961tp,Nambu:1961fr}, is a low energy effective theory
that contains four-point interactions between the quarks. In this
article the following form of the NJL-model is used
\begin{equation}
 \mathcal{L} = \bar \psi \left(i \gamma^\mu \partial_\mu +  \gamma_0 \mu \right) \psi - \mathcal{L}_\mathrm{M}
   + \mathcal{L}_{\bar q q} + \mathcal{L}_\mathrm{det},
\end{equation}
where the mass term of the Lagrangian is 
\begin{equation}
 \mathcal{L}_\mathrm{M} = \bar \psi M_0 \psi,
\end{equation}
and $\mu = (\mu_u, \mu_d)$ denotes the chemical potential. 
Furthermore,
\begin{equation}
 \mathcal{L}_{\bar q q} = G_1 \left[ (\bar \psi \lambda_a \psi)^2 +  
 (\bar \psi \lambda_a i \gamma_5 \psi)^2 \right],
\end{equation}
is the attractive part of the $\bar q q$ channel of the Fierz transformed
color current-current interaction~\cite{Buballa:2003qv} and
\begin{eqnarray}
 \mathcal{L}_\mathrm{det} & = & G_2 e^{i \theta} 
\det \left( \bar \psi_R \psi_L \right) + \mathrm{h.c.} \label{det_int},
\end{eqnarray}
is the 't Hooft determinant interaction which depends on the QCD vacuum 
angle $\theta$. Often $G_1$ and $G_2$ are taken equal, which at
  $\theta=0$ means the low energy spectrum consists of $\sigma$
  and $\bm{\pi}$ fields only. We will restrict to the two flavor
case, using $\lambda_a$ with $a=0,...,3$ as generators of U(2). 
We will not take into account diquark interactions, and therefore do not 
consider color superconductivity that is expected to arise at high baryon 
chemical potential and low temperatures. 
We choose an appropriate basis of quark fields, such that the mass-matrix
$M_0$ is diagonal, i.e.,
\begin{equation}
 \left( \begin{array}{cc}
   m_u & 0 \\
   0   & m_d 
        \end{array}
        \right).
\end{equation}

The symmetry structure of the NJL model is very similar to that of QCD. In
the absence of quark masses and the instanton interaction there is a
global SU(3)$_c \times$U(2)$_R \times$U(2)$_L$-symmetry. The instanton 
interaction breaks it to SU(3)$_c
\times$SU(2)$_L\times$SU(2)$_R\times$U(1)$_B$. For nonzero, but 
equal quark masses this symmetry is reduced to SU(3)$_c
\times$SU(2)$_V\times$U(1)$_B$.  
For unequal quark masses and chemical potentials one is left with 
SU(3)$_c\times$U(1)$_B\times$U(1)$_I$, where $B$ and $I$ stand respectively
for baryon number and isospin.

\section{Chiral transformations and negative quark mass}
It is well known that a theory with $\theta = \pi$ can be related to a
theory with a negative quark mass. Since this sometimes leads to 
confusion concerning the terminology used for the meson spectrum, we will 
elaborate on this relation in this section. 

\subsection{QCD}

We start with the QCD partition function including the $\theta$ term:
\begin{equation}
 Z = \int \uD \psi \uD \bar \psi \uD A \; e^{i \int \ud^4 x\;
   [\mathcal{L}_\mathrm{QCD} +\mathcal{L}_\theta]},
\end{equation}
where 
\begin{eqnarray}
 \mathcal{L}_\mathrm{QCD} & = & \bar \psi \left( i \Dslash - m \right) \psi-
 \frac{1}{4} \Tr F^{\mu \nu} F_{\mu \nu}, \nonumber \\
\mathcal{L}_\theta & = & \frac{\theta g^2}{32 \pi^2} \Tr F^{\mu \nu}
\tilde F_{\mu \nu}. \label{QCDLagrangian}
\end{eqnarray}
The fermion measure of gauge theories is not invariant under chiral 
transformations~\cite{Fujikawa:1979ay}, which can be used to remove 
$\mathcal{L}_\theta$. Since the mass term is not invariant under 
chiral transformations, a $\theta$ dependence then appears in the mass term. 
One obtains
\begin{equation}
 Z =  \int \uD \psi' \uD \bar \psi' \uD A \; e^{i \int \ud^4 x 
\mathcal{L'}_\mathrm{QCD}},
\end{equation}
where the $\theta$-dependence resides in the mass term. Although the
physical results one obtains using the transformed expression will be 
equivalent, one has to be careful when evaluating vacuum expectation
values. 
We define vacuum expectation values of an operator ${\cal O}= {\cal
   O}(\psi,\bar \psi)$ in terms of the original fields (the one of the
   Lagrangian (\ref{QCDLagrangian})) and in terms of the transformed 
or ``primed'' fields as follows:
\begin{eqnarray}
 \braket{{\cal O}}_\theta   & = \int \uD \psi \uD \bar \psi \uD A \; {\cal
   O}(\psi,\bar \psi) \; e^{i \int \ud^4 x [\mathcal{L}_\mathrm{QCD} + 
\mathcal{L}_\theta]}, \nonumber \\
 \braket{{\cal O}'}_\theta & = \int \uD \psi' \uD \bar \psi' \uD A \; 
{\cal O}'(\psi',\bar \psi') \; e^{i \int \ud^4 x \mathcal{L'}_\mathrm{QCD}}.
\end{eqnarray}
Clearly, the condensates $\braket{{\cal O}}_\theta$ and $\braket{{\cal
    O}'}_\theta$ differ for $\theta \neq 0$ and are related by a 
$\theta$-dependent transformation. For instance, 
\begin{equation}
 \braket{\bar \psi \psi}_\theta  = \int \uD \psi \uD \bar \psi \uD A
 \; \bar \psi \psi \; e^{i \int \ud^4 x [\mathcal{L}_\mathrm{QCD} + 
\mathcal{L}_\theta]} \neq \int \uD \psi' \uD \bar \psi' \uD A \; 
\bar \psi' \psi' \; e^{i \int \ud^4 x \mathcal{L'}_\mathrm{QCD}} = 
 \braket{\bar \psi' \psi'}_\theta.
\end{equation}
When discussing a vacuum expectation value like $\braket{\bar \psi
  \psi}_\theta$ it has to be accompanied by a statement about which 
  Lagrangian one is using. 

In what follows we will select the chiral transformation that only
affects the up quark:  
\begin{eqnarray}
  u_L & = e^{-i \theta / 2} u'_L, \nonumber \\
  u_R & = e^{i \theta / 2} u'_R. \label{chiral_transformation}
\end{eqnarray}
This removes $\mathcal{L}_\theta$ from the Lagrangian and the up-quark
mass term changes according to
\begin{equation}
 \bar u m_u u = \bar u' \left[ m_u \cos \theta + m_u i \gamma_5 \sin \theta
 \right] u'.   
\end{equation}
For $\theta = \pi$ a negative up-quark mass results. In addition, 
\begin{equation}
 \braket{\bar u u}_\theta =  \braket{\bar u' u'}_\theta \cos \theta + \braket{\bar u' i \gamma_5 u'}_\theta \sin \theta.
\end{equation}
Below we will use the following notation for the meson condensates:
\begin{align}
  \braket{\sigma} & = \braket{\bar \psi \lambda_0 \psi}, & \braket{\bm{a}_0} &  = \braket{\bar \psi \bm{\lambda} \psi} \nonumber \\
  \braket{\eta} & = \braket{\bar \psi \lambda_0 i \gamma_5 \psi}, & \braket{\bm{\pi}}&  = \braket{\bar \psi \bm{\lambda} i \gamma_5 \psi}.
\end{align}
These condensates transform according to:
\begin{eqnarray}
 \braket{\sigma} & = & \frac{1}{2} \left( \cos \theta + 1 \right) \, \condsigmap + \frac{1}{2} \left( \cos \theta - 1 \right) \, \condazzp +
        \frac{1}{2} \sin \theta \, \condetap + \frac{1}{2} \sin \theta \, \condpizp, \nonumber \\
 \braket{a_0^\pm} & = & \cos \frac{\theta}{2} \, \condazpmp + \sin\frac{\theta}{2} \, \condpipmp, \nonumber \\
 \braket{a_0^0} & = & \frac{1}{2} \left( \cos \theta - 1 \right) \, \condsigmap + \frac{1}{2} \left( \cos \theta + 1 \right) \, \condazzp +
        \frac{1}{2} \sin \theta \, \condetap + \frac{1}{2} \sin \theta \, \condpizp, \nonumber \\
 \braket{\eta} & = & \frac{1}{2} \left( \cos \theta + 1 \right) \, \condetap + \frac{1}{2} \left( \cos \theta - 1 \right) \, \condpizp -
        \frac{1}{2} \sin \theta \, \condsigmap - \frac{1}{2} \sin \theta \, \condazzp, \nonumber \\
 \braket{\pi^\pm} & = & \cos \frac{\theta}{2} \, \condpipmp - \sin\frac{\theta}{2} \, \condazpmp, \nonumber \\
 \braket{\pi^0} & = & \frac{1}{2} \left( \cos \theta - 1 \right) \, \condetap + \frac{1}{2} \left( \cos \theta + 1 \right) \, \condpizp -
        \frac{1}{2} \sin \theta \, \condsigmap - \frac{1}{2} \sin \theta \, \condazzp. \label{trans_condensates}
\end{eqnarray}
Therefore, one has to be careful assigning the names $\pi^0$ and
$\eta$ to the condensates after doing a chiral transformation.  For
example, Ref.~\cite{fujihara:2005wk} discusses a
$\braket{\pi^0}$-condensate using a Lagrangian without $\theta$ term,
but with negative up or down quark mass. This corresponds to an
$\braket{\eta}$-condensate using a Lagrangian with positive quark
masses and a $\theta$-term with $\theta=\pi$. We emphasize that these
  transformations are just a matter of consistently naming mesons and vev's, 
but this is nevertheless important for the comparison of quantities from 
different calculations. 

\subsection{NJL-model}
As the NJL-model is not a gauge-theory, the fermion measure is
invariant under chiral transformations. But now the Lagrangian contains two 
terms that are not invariant under chiral
transformations, the mass-term and the determinant interaction. The latter 
is $\theta$-dependent. Like for QCD, this $\theta$-dependence can be put 
in the up-quark mass term using a chiral transformation. 
So the analysis for the
NJL-model is similar to the analysis for QCD, but instead of a
noninvariant measure we have a noninvariant effective interaction.

The calculation of the ground-state of the NJL-model is more conveniently done 
with the $\theta$-dependence in the up-quark mass term, i.e.\ we use 
\begin{eqnarray}
 \mathcal{L'}_\mathrm{M} & = & \bar u'_R m_u e^{-i \theta} u'_L + \bar d'_R m_d d'_L + \mathrm{h.c.}, \nonumber \\
 \mathcal{L'}_\mathrm{det} & = & G_2 \det \left( \bar \psi'_R \psi'_L \right) + \mathrm{h.c.}
\end{eqnarray}
Therefore, below we will calculate
the effective potential using the transformed (primed) fields, but
discuss the ground state phase structure solely in terms of the
condensates in terms of the original fields. Only in the latter case
the SU(2)$_V$ symmetry 
among the three pions (and among the $a_0$-mesons) is manifest when we
consider degenerate quark masses for instance.

Because we want to investigate the effects of instantons on the
vacuum, we are interested in the dependence on the strength of the 
determinant interaction, which is the effective instanton interaction.
Frank {\it et al.}~\cite{Frank:2003ve} have investigated the effects of 
this interaction at $\theta = 0$, in particular flavor-mixing effects,
on the QCD phase diagram, by choosing the following expressions for $G_1$
and $G_2$ (where our $c$ is their $\alpha$)
\begin{equation}
 G_1 = (1 - c) G_0, \quad G_2 = c G_0. \label{values_G1_G2}
\end{equation}
In this way, the strength of the instanton interaction is controlled
by the parameter $c$, while the value for the quark
condensate at $\theta = 0$ (which is determined by the
  combination $G_1+G_2$) is kept fixed. As mentioned,
  for $G_1=G_2$, or equivalently $c=\half$, only the $\sigma$ and
  $\bm{\pi}$ mesons are present.
For our numerical studies we will use the following values for the
parameters unless stated otherwise: 
$m_u = m_d = 6$ MeV in case of degenerate quark masses, 
a three-dimensional momentum cut-off
$\Lambda = 590$ MeV/$c$ and $G_0 \Lambda^2 = 2.435$. This corresponds
\cite{Frank:2003ve} to a
pion mass of $140.2$ MeV, a pion decay constant
of $92.6$ MeV and finally, a
quark condensate $\braket{\bar u u} = \braket{\bar d d} = (-241.5\ \mathrm{MeV})^3$. These
values are in reasonable agreement with experimental determinations.

\section{Calculation of the ground-state}
To obtain the ground-state of the theory, we 
introduce 8 real condensates as follows
\begin{eqnarray}
 \alpha'_0 & = & - 2 (G_1 + G_2) \condsigmap, \nonumber \\
 \bm{\alpha}' & = & - 2 (G_1 - G_2) \condazp, \nonumber \\
 \beta'_0  & = & - 2 (G_1 - G_2) \condetap, \nonumber \\
 \bm{\beta}'  & = & - 2 (G_1 + G_2) \condpip.
\end{eqnarray}
All quantities in this section refer to the primed fields, but for
notational convenience we will drop the primes from now on in this
section only. Results
presented in the subsequent sections will refer exclusively to the unprimed 
quantities. 

It will be assumed that all condensates are space-time independent. 
A Hubbard-Stratonovich transformation eliminates the four-point
quark interactions, such that the Lagrangian becomes
quadratic in the quark fields and the integration over these fields is
straightforward to perform. 
One obtains the following expression for the thermal effective potential
$\mathcal{V}$ in the mean-field approximation \cite{Warringa:2005jh} 
\begin{equation}
 \mathcal{V} = \frac{\alpha_0^2 + \beta_i^2}{4(G_1 + G_2)} + \frac{\alpha_i^2 + \beta_0^2}{4(G_1 - G_2)} - T \sum_{p_0=(2n+1)\pi T} \int \frac{\ud^3 p}{(2 \pi)^3} \log \det K \label{effpot}
\end{equation}
where $K$ is a matrix in flavor and Dirac space,
\begin{equation}
 K = \mathbbm{1}_\mathrm{f} \otimes (i\gamma_0 p_0 + \gamma_i p_i) -\mu \otimes \gamma_0 - M
\end{equation}
is the inverse quark propagator, and 
\begin{eqnarray}
 M & = & m_u (\cos \theta \, \lambda_u \otimes \mathbbm{1}_\mathrm{d} + \sin \theta \, \lambda_u \otimes i \gamma_5 ) +
 m_d \lambda_d \otimes \mathbbm{1}_\mathrm{d} + \alpha_a \lambda_a \otimes \mathbbm{1}_\mathrm{d}
 + \beta_a \lambda_a \otimes i \gamma_5,\label{mass_matrix}
\end{eqnarray}
with $\lambda_u = (\lambda_1 + \lambda_3)/2$ and $\lambda_d =
(\lambda_1 - \lambda_3)/2$.

The values of the condensates are found by minimizing the effective
potential with respect to these condensates. By exploiting U(1) flavor
symmetry one only has to study the condensates $\alpha_0, \alpha_1,
\alpha_3, \beta_0, \beta_1$, and $\beta_3$.  In
Ref.~\cite{Warringa:2005jh} the $\beta_0$ and $\beta_3$ condensates have been
ignored based on the Vafa-Witten theorem. As we wish to check the
validity of this theorem at finite temperature and density in our
model calculation, we do take these condensates into account.

In order to calculate the effective potential, it is convenient to
multiply $K$ with $\mathbbm{1}_\mathrm{f} \otimes
  \gamma_0$ which
leaves the determinant invariant and yields a new matrix $\tilde{K}$ with $i
p_0$'s on the diagonal. The determinant of $K$ can be calculated as $\det K =
\prod_{i=1}^{8} \left(\lambda_i - i p_0 \right)$, where $\lambda_i$
are the eigenvalues of $\tilde{K}$ with $p_0 = 0$. 
After performing the sum over the Matsubara frequencies, we obtain
\begin{equation}
 T \sum_{p_0 = (2n + 1) \pi T} \log \det K = \sum_{i=1}^{8} \left[ \frac{|\lambda_i|}{2} + T \log \left(1 + e^{-|\lambda_i|/T} \right) \right].
\end{equation}
Finally we need to integrate over the three-momenta $p$ up to the
ultraviolet cutoff $\Lambda$ to determine the effective potential.

Minimizing $\mathcal{V}$ implies solving the equations 
\begin{equation}
 \frac{\partial \mathcal{V}}{\partial x_i} = 0,\label{deriv_eq}
\end{equation}
where $x = \left\{ \alpha_0, \alpha_1, \alpha_3, \beta_0, \beta_1, \beta_3 \right\}$.
The derivatives of the effective potential can be calculated from the 
expression~\cite{Warringa:2005my}
\begin{equation}
 T \frac{\partial}{\partial x_j} \sum_{p_0=(2n + 1) \pi T} \log \det K = \frac{1}{2} \sum_{i=1}^{8} b_{ij} \left(1 - \frac{2}{e^{|\lambda_i|/T} + 1} \right) \mathrm{sgn}\left(\lambda_i \right),
\end{equation}
where $b_{ij} = \left( U^\dagger \partial \tilde{K}(p_0 = 0) /
  \partial x_j U \right)_{ii}$.  Here $U$ is a unitary matrix which
contains in the $i$-th column the normalized eigenvector of
$\tilde{K}$ with eigenvalue $\lambda_i$. Again, one has to integrate
over $p$ to obtain the complete derivative. Since this calculation
does not use the finite distance method, the derivatives can be
determined very accurately. Also, it is very efficient as one needs the
eigenvalues of $\tilde{K}$ anyway in order to calculate the effective
potential.

When a solution to Eq.~\eqref{deriv_eq} has been found, it has to be checked
whether the solution is indeed a minimum and not a maximum or saddle-point.
This is checked by verifying that the Hessian of the solution only has
positive eigenvalues. If more than one minimum is found, the one with the
lowest value is chosen.  Also the continuity of the effective potential is
checked.

The speed of the calculation mainly depends on how fast the
eigenvalues of $\tilde{K}$ can be calculated. To speed up the
evaluation of the calculation of the eigenvalues, one can make use of the
fact that the determinant of $\tilde{K}$ is invariant under the interchanging
of rows and columns. This can be used to bring $\tilde{K}$ to a
block-diagonal form of two $4 \times 4$-matrices. This reduces the
computing time to determine the eigenvalues with a factor of four
as the time to numerically calculate the eigenvalues scales cubically
with the dimension of the matrix.

Another way of improving the speed of the calculation is to choose 
$\vec p$ to lie along the $z$-direction, exploiting the fact that 
$\det \tilde{K}$ does not depend on the direction of $\vec p$.

One final remark we have to make regarding Eq.~\eqref{effpot} is the fact that
in order for the effective potential to have a minimum at finite values for
the condensates, the coupling $G_2$ has to satisfy $- G_1 \leq G_2 \leq G_1$,
and correspondingly, $-\frac{1}{2} \leq c \leq \frac{1}{2}$. From
Eq.~\eqref{det_int} we can see that a negative value for $G_2$ corresponds to
shifting $\theta \to \theta+\pi$, implying that the minimum of the theory will
be at $\theta = \pi$, in violation of the Vafa-Witten theorem at zero
temperature and density.  Therefore, we will restrict to $0 \leq c \leq
\frac{1}{2}$. The case $c = \frac{1}{2}$ is special, because then only the
$\sigma'$ and $\bm{\pi}'$ fields are present in the theory, which means at
$\theta =0$ the $\sigma$ and $\bm{\pi}$ mesons and at $\theta=\pi$ the $\eta$
and $\bm{a}_0$ mesons.

\section{The ground-state of the NJL model}
\begin{figure}[htb]
  \includegraphics[width=0.45\textwidth]{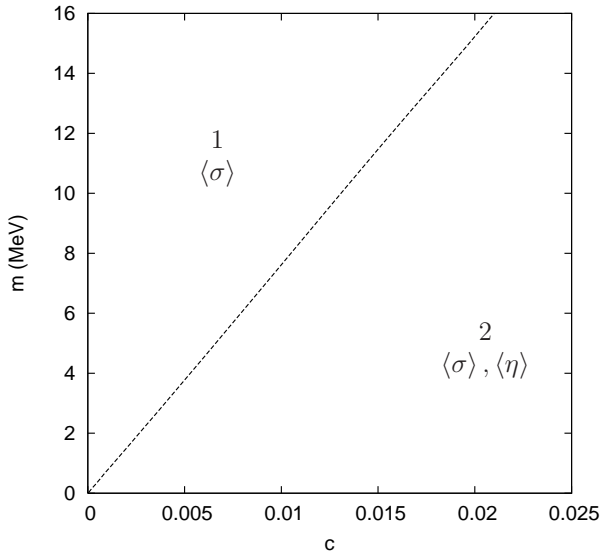}
  \caption{The $(c, m)$ phase diagram at $\theta=\pi$.}
  \label{phasediagram_m_c}
\end{figure}

This section deals with our results for the ground-state of the NJL
model. First we discuss the $\theta$-dependence of the condensates and
the effective potential. It turns out that for nonzero $c$, two
different situations can be distinguished: below a certain critical
$c$ value ($c_\mathrm{crit}$) no spontaneous CP violation takes place
at $\theta = \pi$, whereas for $c$ larger than this critical value
it does take place. The value of this
$c_\mathrm{crit}$ depends on the values of the quark masses. In
Fig.~\ref{phasediagram_m_c} we show the phase diagram at $\theta=\pi$
in the $(c, m)$-plane for degenerate quark masses $m_u = m_d = m$, 
two phases can be distinguished,
\begin{enumerate}
  \item $\braket{\sigma} \neq 0$, the ordinary chiral condensate.
  \item $\braket{\sigma} \neq 0, \braket{\eta} \neq 0$, the CP-violating phase.
\end{enumerate}
The phase transition corresponds to $c_\mathrm{crit}$ and is of second
order. A linear relation exists between the quark mass and
$c_\mathrm{crit}$.

\subsection{The $\theta$-dependence of the vacuum}
When the determinant interaction is turned off, there is no
$\theta$-dependence. In terms of the unprimed fields, only the
$\braket{\sigma}$ condensate is nonzero.

\begin{figure}[htb]
  \includegraphics[width=0.6\textwidth]{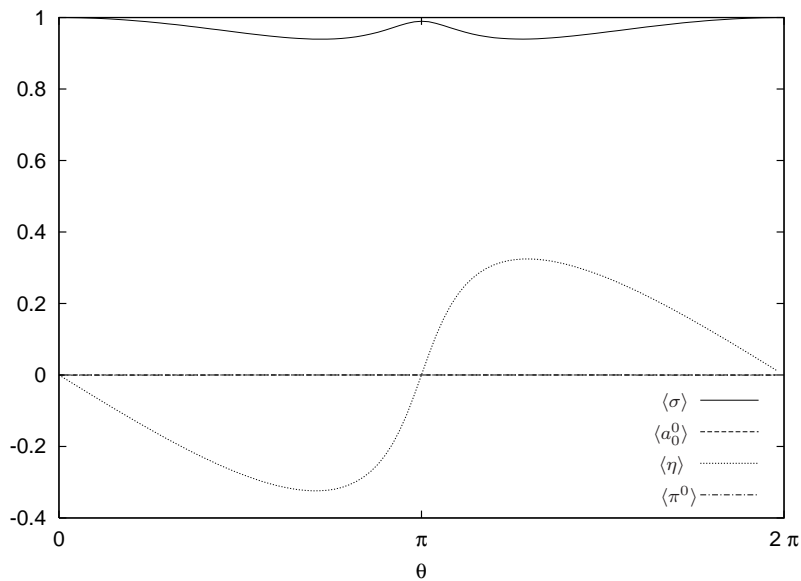}
  \caption{The $\theta$-dependence of the normalized condensates, with
    $c = 0.005 < c_\mathrm{crit}$.}
  \label{vac_cond_0.005}
\end{figure}

In Fig.~\ref{vac_cond_0.005} we show the $\theta$-dependence of the various
condensates for the case $c = 0.005$, which for our choice $m_u = m_d = 6 \,
\mathrm{MeV}$ is below $c_\mathrm{crit} \approx 0.008$. As can be seen, no
spontaneous CP violation occurs, since $\braket{\eta}=0$ at $\theta=\pi$.
Explicit CP violation for other values of $\theta$ does occur, as expected. In
this figure the condensates are normalized with respect to $\braket{\sigma}$
at $\theta = 0$.  Both $\braket{\bm{\pi}}$ and $\braket{\bm{a}_0}$ are zero
for all $\theta$ and this remains true for $c$ above $c_\mathrm{crit}$ for
degenerate quark masses.

\begin{figure}[htb]
  \includegraphics[width=0.6\textwidth]{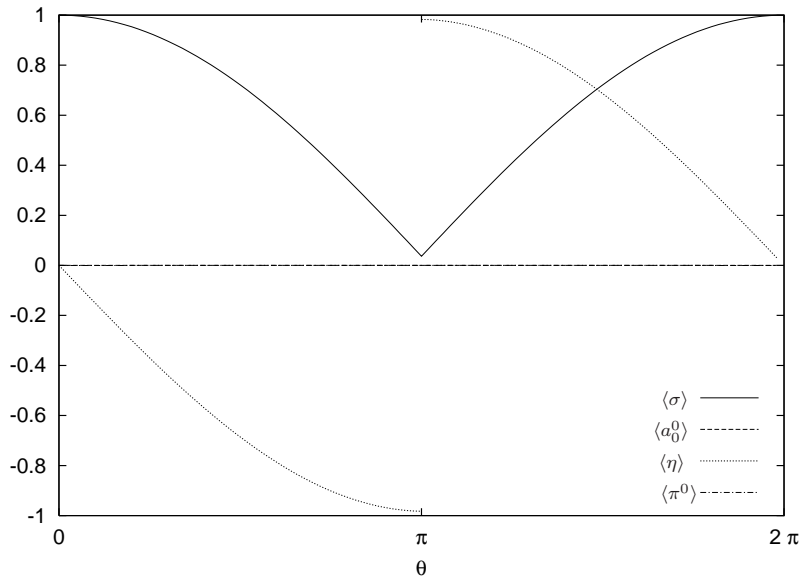}
  \caption{The $\theta$-dependence of the normalized condensates, with
    $c = 0.2 > c_\mathrm{crit}$.}
  \label{vac_cond_0.21}
\end{figure}

\begin{figure}[htb]
  \includegraphics[width=0.6\textwidth]{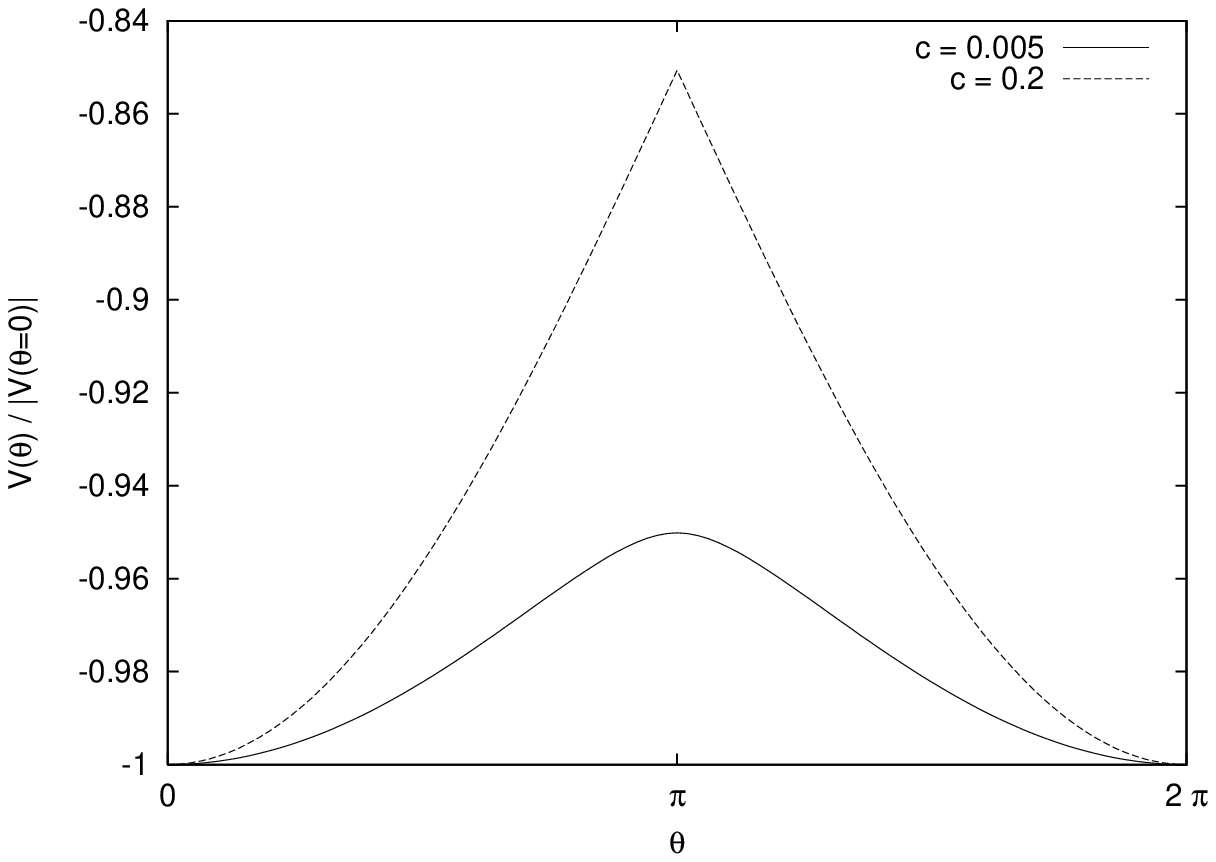}
  \caption{The $\theta$-dependence of the normalized effective
    potential at $c = 0.005$ and $c=0.2$.}
  \label{eff_pot_0.005_0.21}
\end{figure}

Fig.~\ref{vac_cond_0.21} shows the case of $c = 0.2$. Spontaneous CP
violation is clearly visible, as $\braket{\eta}$ is nonzero at
$\theta = \pi$. As can be seen two degenerate vacua then exist, 
with opposite signs for $\braket{\eta}$.
These two degenerate vacua differ by a CP transformation. This is 
known as Dashen's phenomenon~\cite{Dashen:1970et} and is also apparent
from the $\theta$-dependence of the effective potential.
In Fig.~\ref{eff_pot_0.005_0.21} we show the effective potential as a
function of $\theta$ normalized to its value at $\theta=0$, for
the two cases $c = 0.005$ and $c =0.2$. In both cases, the minimum of the
effective potential is at $\theta = 0$, in agreement with
the Vafa-Witten theorem. Furthermore, it can be seen that the case
with spontaneous CP-violation has a cusp at $\theta = \pi$, and
therefore a left and a right derivative which differ by a sign. 
Due to the axial anomaly, the
$\theta$-derivative of the effective potential is proportional to
$\braket{\eta}$. This explains the occurrence of two values for the
$\eta$ condensate.  

\subsection{Phase structure at $\theta = \pi$}
In this section we concentrate further on the case $\theta = \pi$.
We will start with a discussion of the
mass-dependence of the ground-state.  From Ref.~\cite{Creutz:2003xu} we
know that in three-flavor chiral perturbation theory a region exists
in the $(m_u,m_d)$-plane where CP is spontaneously violated, cf.\
Fig.~\ref{ChPT_NF3_mu_md}. In that case 
the shape of the CP violating region depends on the strange quark mass.
In the present case it depends on the choice of $c$. 

\begin{figure}[htb]
  \includegraphics{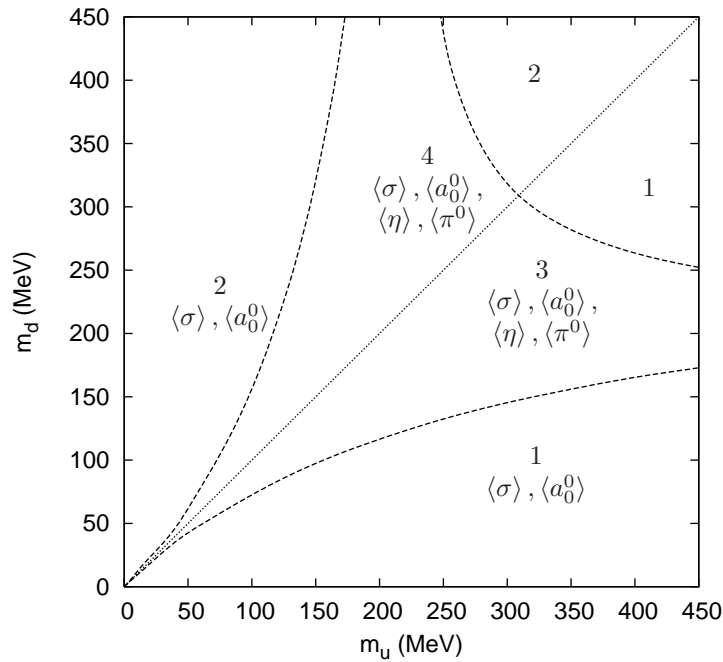}
  \caption{The $(m_u,m_d)$ phase diagram of the NJL at $\theta = \pi$
    with $c =0.4$. The dashed lines denote second order
    phase transitions and the dotted line a crossover.}
  \label{mu_md_phase_diagram}
\end{figure}

In Fig.~\ref{mu_md_phase_diagram} we show the phase diagram of the NJL
model at $\theta = \pi$ with $c = 0.4$ in the $(m_u,m_d)$-plane. Four
phases can be distinguished
\begin{enumerate}
  \item $\braket{\sigma} < 0, \braket{a_0^0} < 0$ 
\label{phase1}
  \item $\braket{\sigma} < 0, \braket{a_0^0} > 0$ 
\label{phase2}
  \item $\braket{\sigma} <0, \braket{a_0^0} < 0, \braket{\eta} \neq 0,
  \braket{\pi^0} \neq 0$ 
\label{phase3}
  \item $\braket{\sigma} < 0, \braket{a_0^0} > 0, \braket{\eta} \neq
  0, \braket{\pi^0} \neq 0$  
\label{phase4}
\end{enumerate}
In phases \ref{phase3} and
\ref{phase4} two degenerate vacua exist with opposite signs for
both $\braket{\eta}$ and $\braket{\pi^0}$.  
The phase transitions between the CP-conserving phases \ref{phase1}
and \ref{phase2} to the CP-violating phases \ref{phase3} and
\ref{phase4} are second order. The phases \ref{phase1} and
\ref{phase2} only differ in the sign for the
$\braket{a_0^0}$-condensate, the same holds for the phases
\ref{phase3} and \ref{phase4}. The phase transition between the phases
\ref{phase3} and \ref{phase4} is a crossover, as is the case for the
phase transition between phase~\ref{phase1} and \ref{phase2} for large
$m_u$ and $m_d$. Exactly at the crossover, $\braket{a_0^0}$ vanishes
and in the CP-violating region the same applies to $\braket{\pi^0}$,
but not to $\braket{\eta}$. 
The fact that $a_0^0$-condensation (and
$\pi^0$-condensation in the CP-violating region) occurs when the
masses are not equal simply reflects the explicit breaking of SU(2)$_V$
which occurs for nondegenerate quark masses. 

The shape of the CP-violating region is determined by the asymptotes, which
are proportional to $c$. We conclude that in contrast to two-flavor
lowest-order chiral perturbation theory (the case of $m_s \to \infty$ in Ref.\
\cite{Creutz:2003xu}, such that the asymptotes are moved to $m_u=m_d=\infty$),
the NJL model does have a spontaneous CP-violating phase for two nondegenerate
quark flavors.  This is in accordance with the chiral
perturbation theory analysis of Ref.\ \cite{Tytgat:1999yx} in the large $N_c$
limit and for finite $a/N \gg m_u,m_d$.

\section{Finite temperature and baryon chemical potential}
\begin{figure}[htb]
  \begin{minipage}[t]{0.45\linewidth}
    \centering
    \includegraphics[width=\textwidth]{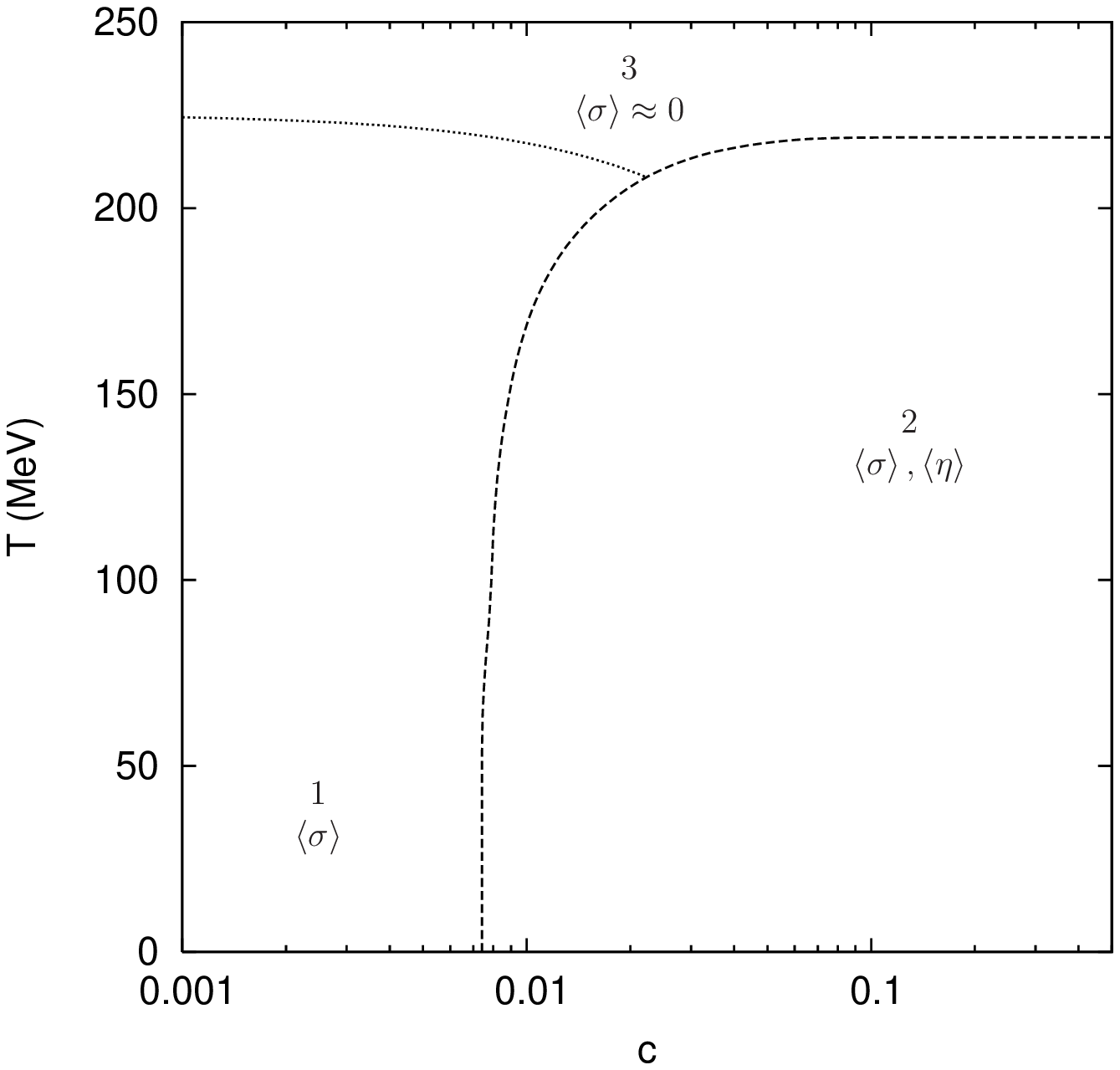}
    \caption{The $(T,c)$ phase diagram of the NJL model at $\theta =
      \pi$. The dotted line represents a crossover, which is defined
      by the inflection points $\partial^2 \braket{\sigma} /
      \partial T^2 = 0$.}
    \label{c_T_phasediagram}
  \end{minipage}%
  \hspace{0.1\linewidth}%
  \begin{minipage}[t]{0.45\linewidth}
    \centering
    \includegraphics[width=\textwidth]{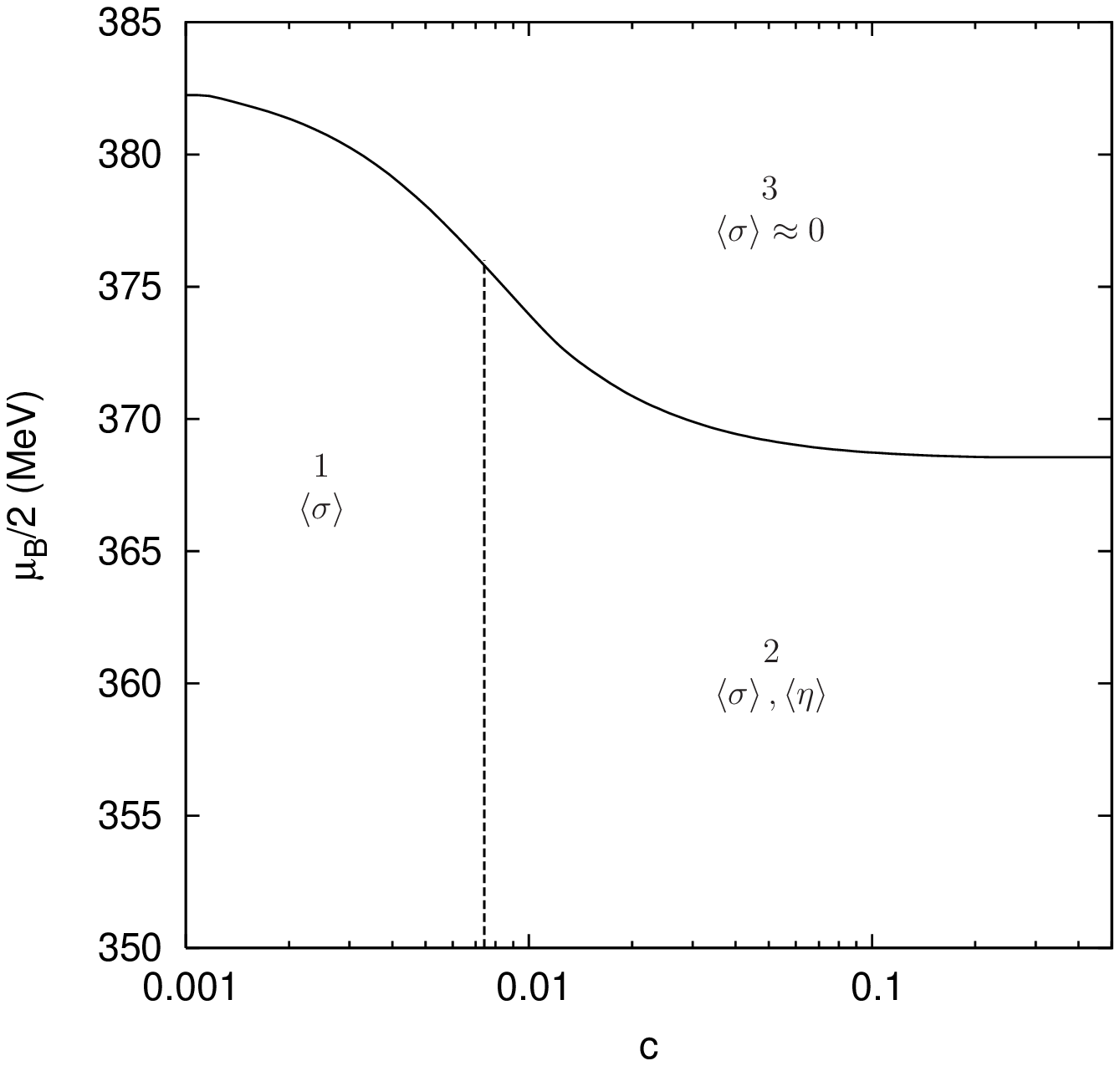}
    \caption{The $(\mu_B,c)$ phase diagram of the NJL model at $\theta =
      \pi$.}
    \label{muB_c_phasediagram}
  \end{minipage}
\end{figure}

In this section we turn to the changes in the phase structure at nonzero
temperature and density.  Ref.~\cite{fujihara:2005wk} states that the
CP-violating phase at $\theta=\pi$ does not exist at high temperatures, i.e.\
a critical temperature exists above which the CP-violating condensates are
zero.  Ref.~\cite{fujihara:2005wk} only considered the case $c = 0.5$. Here we
generalize their results to other $c$ values. In Fig.~\ref{c_T_phasediagram}
the $(T,c)$ phase diagram is shown for degenerate quark masses.
The following three phases arise
\begin{enumerate}
  \item[1.] $\braket{\sigma} \neq 0$, the ordinary chiral condensate.
  \item[2.] $\braket{\sigma} \neq 0, \braket{\eta} \neq 0$, the CP-violating phase.
  \item[3.] $\braket{\sigma} \approx 0$, the (almost) chiral symmetry
  restored phase.
\end{enumerate}
The phase structure at $T=0$ can be understood from Fig.\
\ref{mu_md_phase_diagram}: for degenerate quark masses the two phases are
encountered on its diagonal. The phase transition occurs at that particular
value of $m_u=m_d$ for which $c=0.4$ is the critical $c$. The phase transition
between phases 1 and 2 is of second order for all temperatures.  For
nondegenerate quark masses these two phases would be phases \ref{phase1} and
\ref{phase3} or \ref{phase2} and \ref{phase4} of Fig.\
\ref{mu_md_phase_diagram} depending on whether $m_u$ is larger or smaller than
$m_d$, respectively. In that case two second order phase transitions are
present.

Above a certain temperature one observes in Fig.~\ref{c_T_phasediagram} an
approximate restoration of chiral symmetry (phase~3). Note that the chiral
symmetry is not fully restored due to the quark masses. The phase transition
between phases 1 and 3 is a crossover, like it is at $\theta = 0$. The
crossover line is defined by the inflection points $\partial^2 \braket{\sigma}
/ \partial T^2 = 0$.
 
At high temperature also the CP-violating phase disappears. This is consistent
with the fact that at high temperature instanton effects become exponentially
suppressed \cite{Gross:1980br}. The CP-violating phase is after all realized
due to the instanton induced interaction. The maximum value of the critical
temperature as function of $c$ is 219 MeV.

We have verified that also for nonzero temperature the minimum of the
effective potential is at $\theta=0$, which means the Vafa-Witten theorem
continues to hold in the NJL model at nonzero temperature. The same applies to
baryon and isospin chemical potential. We also have checked whether there are
any local minima in the effective potential at nonzero temperature and
density, but we found none.

Now we will briefly consider nonzero baryon chemical potential
$\mu_B=\mu_u+\mu_d$, where $\mu_{u,d}$ denote the $u,d$ quark chemical
potentials. The $(\mu_B,c)$ phase diagram is displayed in
Fig.~\ref{muB_c_phasediagram} for a restricted range of $\mu_B$ values. The
same phases occur as in the $(T, c)$ phase diagram, but now the phase
transition to the (almost) chiral symmetry restored phase is of first order,
like for $\theta = 0$.  Furthermore, the first-order phase transition has a
small $c$ dependence. As always, the phase transition from
phase~1 to phase~2 is of second order.

\section{Nonzero isospin chemical potential}
In quark matter systems equilibrium and neutrality conditions can require that
$\mu_u \neq \mu_d$.  Son and Stephanov \cite{Son:2000xc} observed that charged
pion condensation can occur for nonzero isospin chemical potential
$\mu_I=\mu_u-\mu_d$. At $\theta = 0$ this second order phase transition
between the ordinary phase of broken chiral symmetry ($\braket{\sigma}\neq 0$)
to the pion condensed phase (which also breaks chiral symmetry) occurs when
$\mu_I$ equals the vacuum pion mass. In this subsection we address this issue
at $\theta=\pi$.

\begin{figure}[htb]
  \includegraphics{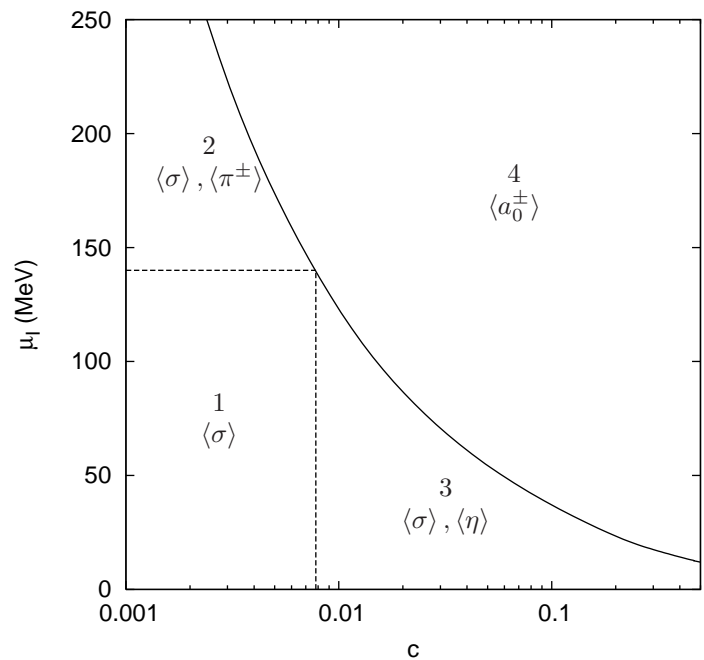}
  \caption{The $(\mu_I,c)$ phase diagram of the NJL model at $\theta =
    \pi$.}
  \label{muI_c_phasediagram}
\end{figure}

In Fig.~\ref{muI_c_phasediagram} we show the phase diagram of the NJL model in
the $(\mu_I,c)$-plane, for $m_u=m_d=6$ MeV.  The solid line indicates a
first-order phase transition, the dashed lines indicate second-order phase
transitions. The four phases are characterized as follows
\begin{enumerate}
  \item $\braket{\sigma} \neq 0$
\label{chiral_phase}
  \item $\braket{\sigma} \neq 0$, $\braket{\pi^\pm} \neq 0$
\label{pion_phase}
  \item $\braket{\sigma} \neq 0$, $\braket{\eta} \neq 0$
\label{eta_phase}
  \item $\braket{a_0^\pm} \neq 0$
\label{a0pm_phase}
\end{enumerate}
Phase \ref{a0pm_phase} is a novel phase characteristic of $\theta =
\pi$. This phase also has a small nonzero $\braket{\sigma}$-condensate
(not indicated), due to the explicit breaking by the quark masses.

For $c < c_\mathrm{crit}$ a nonzero $\braket{\pi^\pm}$-condensate exists above
a certain $\mu_I$ value. Like at $\theta=0$ the second order phase transition
turns out to be at $\mu_I = m_\pi$, where $m_\pi$ is the vacuum pion mass. In
addition, there is a second phase transition, a first-order phase transition,
at larger $\mu_I$, where charged pion condensation makes way for charged $a_0$
condensation.  For $c > c_\mathrm{crit}$ no nonzero
$\braket{\pi^\pm}$-condensate exists, only nonzero $\braket{a_0^\pm}$. The
phase transition between phases \ref{eta_phase} and \ref{a0pm_phase} is of
first-order.  The question arises what determines the value of $\mu_I$ at this
phase transition to charged meson condensation? To answer this question,
the meson masses need to be calculated.

\subsection{The $c$-dependence of the meson masses and mixing}

\begin{figure}[htb]
  \includegraphics{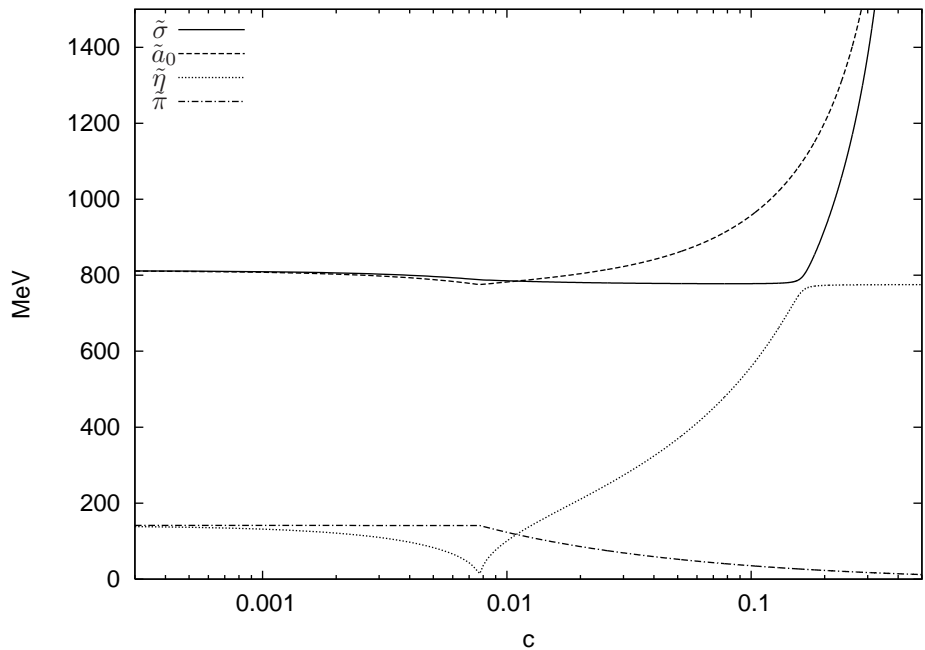}
  \caption{The $c$-dependence of the meson masses at $\theta = \pi$. The masses are calculated in the RPA.}
  \label{c_dependence_masses}
\end{figure}

\begin{figure}[htb]
  \includegraphics[width=0.6\textwidth]{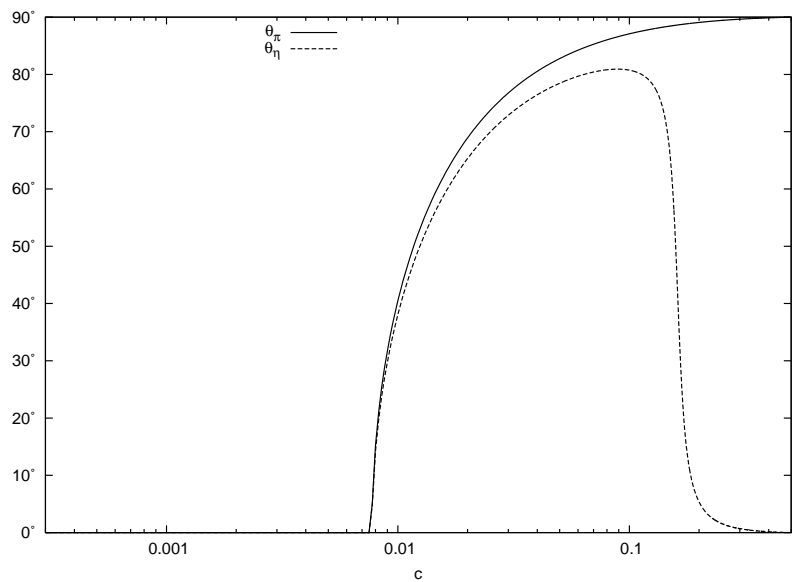}
  \caption{The $c$-dependence of the mixing-angle of the mesons at $\theta = \pi$. The mixing is  calculated in the RPA.}
  \label{c_dependence_mixing}
\end{figure}

As said, at $\theta = 0$ charged pion condensation occurs when $\mu_I$ is
larger or equal to the vacuum ($\mu_I = 0$) pion mass. For the NJL model this
has been studied extensively in
Refs.~\cite{Barducci:2004tt,Barducci:2004nc,He:2005sp,Warringa:2005jh}.  This
condition is independent of $c$. To see what happens in the $\theta = \pi$
case, we calculate the $c$-dependence of the meson masses, with $\mu_I = 0$.
The results are shown in Fig.~\ref{c_dependence_masses} and in
Fig.~\ref{c_dependence_mixing} the $c$-dependence of the mixing is shown.
Clearly, at $\theta=\pi$ the situation quite different from $\theta=0$.


At $c=0$ (no instanton interactions) the $\eta$ and $\bm{\pi}$ masses are
equal and also the $\sigma$ and $\bm{a}_0$ masses. This follows from the
symmetry of the Lagrangian at $c = 0$, which has a U(2)$_L \otimes$
U(2)$_R$-symmetry that is spontaneously broken by the chiral condensate
(ignoring the explicit breaking by the quark masses) to U(2)$_V$. This means
there are four (pseudo-)Goldstone bosons, with the same (small) masses: the
$\eta$ and $\bm{\pi}$ mesons.  The instanton interactions remove the
degeneracy for $c\neq 0$.

When $c > c_\mathrm{crit}$, i.e.\ when $\braket{\eta} \neq 0$, a complication
arises: the mass eigenstates are not CP or P eigenstates any longer. The
occurrence of the $\eta$ condensate results in mixing of the $\sigma$-particle
with its parity partner, the $\eta$-particle. Similarly, the pions mix with
their parity partners, the $\bm{a}_0$'s. The mixing is to be expected because
when the ground-state is not CP-conserving, there is no need for the
excitations, i.e.\ the mesons, to be CP eigenstates or states of definite
parity in case of charged mesons.

The mass eigenstates, denoted with a tilde, are defined in the
following way 
\begin{eqnarray}
  \ket{\tilde \sigma} & = & \cos \theta_\eta \ket{\sigma} + \sin \theta_\eta \ket{\eta}, \nonumber \\
  \ket{\tilde \eta}   & = & \cos \theta_\eta \ket{\eta}   - \sin \theta_\eta \ket{\sigma}, \nonumber \\
  \ket{\tilde{\bm{a}}_0}  & = & \cos \theta_{\bm{\pi}} \ket{\bm{a}_0} + \sin \theta_{\bm{\pi}} \ket{\bm{\pi}}, \nonumber \\
  \ket{\tilde{\bm{\pi}}}   & = & \cos \theta_{\bm{\pi}} \ket{\bm{\pi}} - \sin \theta_{\bm{\pi}} \ket{\bm{a}_0},
\end{eqnarray}
where $\theta_\eta$ and $\theta_{\bm{\pi}}$ are the mixing angles. The states
on the r.h.s.\ are the usual states of definite parity.

The calculation of the mixing and the resulting masses is similar to the
mixing of $\eta_0$ and $\eta_8$ in the three-flavor NJL-model, which was
discussed in great detail in Ref.~\cite{Klevansky:1992qe} using the random
phase approximation (RPA), the approach we will also employ here. As a side
remark, we mention that we have also calculated the curvature of the effective
potential at the minimum which should {\em approximately\/} 
be proportional to
the masses, giving indeed very similar results\footnote{This method was also
  used in Ref.\ \cite{Barducci:2004tt} to calculate the pion mass in order to
  check the condition $\mu_I \geq m_\pi$ for charged pion condensation. A
  discrepancy of approximately 25\% was obtained.}.

When $c < c_\mathrm{crit}$ no mixing takes place and the tilde fields are
equal to their counterparts without tilde. When $c > c_\mathrm{crit}$ mixing
occurs. The mixing between $\eta$ and $\sigma$ increases rapidly as $c$
increases, reaching a maximum at $c = 0.09$, where $\tilde \sigma$ is almost
completely $\eta$ and vice versa. For larger $c$ the mixing however decreases
rapidly again, so that when $c = \half$, $\tilde \sigma$ ($\tilde \eta$) is
again equal to $\sigma$ ($\eta$).

The mixing between $\bm{a}_0$ and the pions behaves differently, here
the mixing angle increases rapidly to become $90^\circ$ at $c=\half$, 
i.e.\ $\tilde{\bm{\pi}}$ becomes $\bm{a}_0$ and vice versa.

Now we turn to the behavior of the tilde-meson masses, which also display
unusual features as function of $c$.  When $c < c_\mathrm{crit}$, the
$\tilde{\bm{\pi}}$ masses are constant, and equal to the ordinary pion masses.
Furthermore, the $\tilde{\eta}$ mass decreases with increasing $c$.  This is
peculiar to $\theta = \pi$, because at $\theta = 0$ the $\eta$ mass increases
with increasing $c$. The $\tilde{\eta}$ mass has its lowest, nonzero value at
$c_{\mathrm{crit}}$. This is in contrast to three-flavor lowest order
  chiral perturbation theory~\cite{Creutz:2003xu}, where the $\eta$ mass (in
  Ref.~\cite{Creutz:2003xu} actually the $\pi^0$ mass using the primed theory)
vanishes at the phase transition. Finally, also the masses of the $\bm{a}_0$'s
and $\sigma$ decrease slightly with increasing $c$.

When $c > c_\mathrm{crit}$ the $c$-dependence of the masses changes
dramatically. The $\tilde{\bm{\pi}}$ mass now decreases
monotonically with increasing $c$, whereas 
the $\tilde{\bm{a}}_0$ and $\tilde{\sigma}$ mass both 
increase monotonically to infinity towards $c = \half$. The latter can
be understood, because when $c = \half$, $G_1$ equals $G_2$, which
as mentioned means for $\theta = \pi$ that there are no $\bm{\pi}$ and
$\sigma$ mesons in the spectrum.  

Another striking feature is that the $\tilde{\eta}$ mass rises until it almost
reaches the $\tilde{\sigma}$ mass, after which it remains approximately
constant. The behavior of the $\tilde{\sigma}$ mass is opposite, first it is
almost constant and when it becomes almost equal to the $m_{\tilde{\eta}}$
mass it increased to infinity.  The masses of $\tilde{\sigma}$ and
$\tilde{\eta}$ cannot cross when there are interactions that mix the two
states, which is similar to level repulsion in quantum mechanics. The point
where both masses are almost equal corresponds to a mass that is twice the
constituent quark mass. This forms the threshold to decay into two quarks,
which makes one of the two mesons unstable when $c> c_\mathrm{crit}$.

Now we turn again to the original question concerning the charged meson
condensation phase transition. From the calculation of the masses of the
tilde-mesons, we infer that the condition for charged meson condensation at
$\theta=\pi$ is $\mu_I \geq m_{\tilde{\pi}}(c)$.  For $c < c_{\mathrm{crit}}$,
the phase transition takes place when $\mu_I$ equals $m_{\tilde{\pi}}
=m_{\pi}$, as it does at $\theta=0$.  For $c > c_\mathrm{crit}$ it takes place
at the mass of $\tilde{\bm{\pi}}$, which is now a mixed state of $\bm{\pi}$
and $\bm{a}_0$. At $c=\half$ this means at the mass of the $\bm{a}_0$. The
latter observation is in agreement with a result of
Ref.~\cite{Metlitski:2005di}, where the $(\mu_I,\theta)$ phase diagram of
degenerate two-flavor chiral perturbation theory is investigated to lowest
order at effectively $c=\half$ (due to the absence of the $\eta$ meson). There
it is observed that charged pion condensation occurs when $\mu_I$ is equal
to the $\theta$-dependent pion mass $m_\pi(\theta)$.  In
Ref.~\cite{Metlitski:2005di} all $\theta$-dependence resides in the mass
matrix, with both quarks having a $\theta$-dependent mass. Hence, their
$\theta$-dependent pion field corresponds to what we call $\pi'$, which at
$\theta=\pi$ is the $a_0$ field in the original, unprimed theory, leading to
an agreement with our finding.

The second phase transition for $c < c_\mathrm{crit}$ from the charged pion
condensed phase to the charged $a_0$ condensed phase does not correspond to
$\mu_I$ being equal to a meson mass calculated at $\mu_I=0$. Although we have
not found a condition in terms of the calculated masses, $\mu_I$ at this
first-order phase transition follows a line that is the smooth continuation of
the $\tilde{\pi}$ mass in the region $c > c_\mathrm{crit}$ to infinity at
  $c=0$. A calculation of the meson masses at nonzero $\mu_I$, such as
  performed in Ref.\ \cite{He:2005sp}, might resolve this open issue.

\section{Conclusion and discussion}
The $\theta$-dependence of the ground-state of the two-flavor NJL model is
investigated in the mean-field approximation. The main focus is on the case
$\theta=\pi$, when spontaneous CP-violation is possible. The
$\theta$-dependence of the theory is found to strongly depend on the strength
of the 't Hooft determinant interaction. When the strength of this
interaction, which is governed by the parameter $c$, is small or zero, no
spontaneous CP-violation takes place at $\theta = \pi$. The low-energy physics
is then almost the same as at $\theta = 0$. At larger $c$ however, spontaneous
CP-violation does take place at $\theta = \pi$. So the phenomenon of
spontaneous CP-violation is governed by the 't Hooft determinant interaction,
which describes the effect of instantons in the effective theory. The question
whether $c$ is sufficiently large for CP-violation to occur at $\theta = \pi$
depends on the quark masses. In other words,
spontaneous CP-violation requires instantons, but its actual realization
depends on the size of their contribution w.r.t.\ the quark masses. This is
also expected to be the case in QCD, where it can be phrased in terms of the
low-energy theorem identity $\sum_q 2i m_q \braket{\bar{q}\gamma_5 q} = - N_f
\braket{g^2 F \tilde{F}}/8 \pi^2$ (cf.\ e.g.\ Ref.~\cite{Metlitski:2005db}),
which relates $\braket{\eta}$ to the first derivative of the effective
potential w.r.t.\ $\theta$. Depending on $m_q$ the coupling constant $g$ needs
to be sufficiently large for spontaneous CP-violation to take place. Or in
other words, the energy needs to be sufficiently low.  The latter observation
is in agreement with the disappearance of the CP-violation at temperatures
above a certain critical temperature or density. Therefore, we conclude that
spontaneous CP-violation in the strong interaction is an inherently low-energy
phenomenon.

We have checked that the Vafa-Witten theorem holds in the NJL model also at
finite temperature and density and found that no local minima arise,
  indicating the absence of meta-stable CP-violating states in the NJL model.
We have confirmed several previous results that were obtained in two-flavor
chiral perturbation theory.  We found (in accordance with the results of Ref.\
\cite{Tytgat:1999yx}) that two-flavor lowest-order chiral perturbation theory
with $a/N \to \infty$ is in general not rich enough to yield results 
that one might expect to hold in QCD too. 
It leads for instance to the conclusion that only for $m_u=m_d$
spontaneous CP violation occurs, without a critical strength of the 
instanton induced interaction. In contrast, the phase diagram of the
two-flavor NJL model is very similar to that of three-flavor chiral
perturbation theory~\cite{Creutz:2003xu}, where spontaneous CP violation 
arises for specific ranges of quark masses.

We also found that the presence of a nonzero $\eta$-condensate has a strong
effect on the $c$-dependence of the meson masses and gives rise to mixing
among the states of definite parity, as expected when CP invariance is not a
symmetry anymore.  As a result, the pions mix with their parity partners, the
$\bm{a}_0 $'s, and the $\eta$ meson mixes with its parity partner, the
$\sigma$ meson. Unlike the mixing discussed as a function of $\theta$ which is
just a matter of consistently naming the states in order to be able to compare
to results obtained with negative quark masses and which does not affect
physical results, the mixing as function of $c$ does change the physics. For
instance, the condition for charged pion condensation at nonzero isospin
chemical potential becomes modified.  At $\theta=\pi$ for $c <
c_{\mathrm{crit}}$, a second-order phase transition takes place when $\mu_I$
equals $m_{\pi}$, just as at $\theta=0$ found by Son and Stephanov.  However,
we find that for $c > c_\mathrm{crit}$ it becomes a first-order phase
transition to a novel phase of charged $a_0$ condensation that takes
place at the mass of $\tilde{\bm{\pi}}$, which is a mixed state of $\bm{\pi}$
and $\bm{a}_0$. At $c=\half$ it is entirely $\bm{a}_0$. Charged $a_0$
condensation also arises for $c < c_{\mathrm{crit}}$ and $\mu_I > m_{\pi}$,
but it appears there is no condition in terms of vacuum meson masses for this
second phase transition.

We expect the presented two-flavor NJL model results to remain valid in the
case of three flavors and when going beyond the mean-field approximation, but
this remains to be studied. It would be very interesting if the results could
in the future be compared to lattice QCD results on the low-energy physics at
$\theta=\pi$.

\begin{acknowledgments}
We would like to thank Harmen Warringa for kindly sharing with us his code
to calculate the effective potential. We also thank the members of the
theory group at the VU, in particular Wilco den Dunnen and Erik
Wessels, for fruitful discussions.
\end{acknowledgments}

\bibliography{references}

\begin{thebibliography}{42}
\expandafter\ifx\csname natexlab\endcsname\relax\def\natexlab#1{#1}\fi
\expandafter\ifx\csname bibnamefont\endcsname\relax
  \def\bibnamefont#1{#1}\fi
\expandafter\ifx\csname bibfnamefont\endcsname\relax
  \def\bibfnamefont#1{#1}\fi
\expandafter\ifx\csname citenamefont\endcsname\relax
  \def\citenamefont#1{#1}\fi
\expandafter\ifx\csname url\endcsname\relax
  \def\url#1{\texttt{#1}}\fi
\expandafter\ifx\csname urlprefix\endcsname\relax\def\urlprefix{URL }\fi
\providecommand{\bibinfo}[2]{#2}
\providecommand{\eprint}[2][]{\url{#2}}

\bibitem[{\citenamefont{Baluni}(1979)}]{Baluni:1978rf}
\bibinfo{author}{\bibfnamefont{V.}~\bibnamefont{Baluni}},
  \bibinfo{journal}{Phys. Rev. D} \textbf{\bibinfo{volume}{19}},
  \bibinfo{pages}{2227} (\bibinfo{year}{1979}).

\bibitem[{\citenamefont{Crewther et~al.}(1979)\citenamefont{Crewther,
  di~Vecchia, Veneziano, and Witten}}]{Crewther:1979pi}
\bibinfo{author}{\bibfnamefont{R.}~\bibnamefont{Crewther}},
  \bibinfo{author}{\bibfnamefont{P.}~\bibnamefont{di~Vecchia}},
  \bibinfo{author}{\bibfnamefont{G.}~\bibnamefont{Veneziano}},
  \bibnamefont{and} \bibinfo{author}{\bibfnamefont{E.}~\bibnamefont{Witten}},
  \bibinfo{journal}{Phys. Lett.} \textbf{\bibinfo{volume}{88B}},
  \bibinfo{pages}{123} (\bibinfo{year}{1979}).

\bibitem[{\citenamefont{Kawarabayashi and Ohta}(1981)}]{Kawarabayashi:1980uh}
\bibinfo{author}{\bibfnamefont{K.}~\bibnamefont{Kawarabayashi}}
  \bibnamefont{and} \bibinfo{author}{\bibfnamefont{N.}~\bibnamefont{Ohta}},
  \bibinfo{journal}{Prog. Theor. Phys.} \textbf{\bibinfo{volume}{66}},
  \bibinfo{pages}{1789} (\bibinfo{year}{1981}).

\bibitem[{\citenamefont{Vafa and Witten}(1984)}]{Vafa:1984xg}
\bibinfo{author}{\bibfnamefont{C.}~\bibnamefont{Vafa}} \bibnamefont{and}
  \bibinfo{author}{\bibfnamefont{E.}~\bibnamefont{Witten}},
  \bibinfo{journal}{Phys. Rev. Lett.} \textbf{\bibinfo{volume}{53}},
  \bibinfo{pages}{535} (\bibinfo{year}{1984}).

\bibitem[{\citenamefont{Dashen}(1971)}]{Dashen:1970et}
\bibinfo{author}{\bibfnamefont{R.}~\bibnamefont{Dashen}},
  \bibinfo{journal}{Phys. Rev. D} \textbf{\bibinfo{volume}{3}},
  \bibinfo{pages}{1879} (\bibinfo{year}{1971}).

\bibitem[{\citenamefont{Witten}(1980)}]{Witten:1980sp}
\bibinfo{author}{\bibfnamefont{E.}~\bibnamefont{Witten}},
  \bibinfo{journal}{Ann. Phys.} \textbf{\bibinfo{volume}{128}},
  \bibinfo{pages}{363} (\bibinfo{year}{1980}).

\bibitem[{\citenamefont{di~Vecchia and Veneziano}(1980)}]{DiVecchia:1980ve}
\bibinfo{author}{\bibfnamefont{P.}~\bibnamefont{di~Vecchia}} \bibnamefont{and}
  \bibinfo{author}{\bibfnamefont{G.}~\bibnamefont{Veneziano}},
  \bibinfo{journal}{Nucl. Phys. B} \textbf{\bibinfo{volume}{171}},
  \bibinfo{pages}{253} (\bibinfo{year}{1980}).

\bibitem[{\citenamefont{Smilga}(1999)}]{Smilga:1998dh}
\bibinfo{author}{\bibfnamefont{A.~V.} \bibnamefont{Smilga}},
  \bibinfo{journal}{Phys. Rev. D} \textbf{\bibinfo{volume}{59}},
  \bibinfo{pages}{114021} (\bibinfo{year}{1999}).

\bibitem[{\citenamefont{Tytgat}(2000)}]{Tytgat:1999yx}
\bibinfo{author}{\bibfnamefont{M.~H.~G.} \bibnamefont{Tytgat}},
  \bibinfo{journal}{Phys. Rev. D} \textbf{\bibinfo{volume}{61}},
  \bibinfo{pages}{114009} (\bibinfo{year}{2000}).

\bibitem[{\citenamefont{Akemann et~al.}(2002)\citenamefont{Akemann, Lenaghan,
  and Splittorff}}]{Akemann:2001ir}
\bibinfo{author}{\bibfnamefont{G.}~\bibnamefont{Akemann}},
  \bibinfo{author}{\bibfnamefont{J.~T.} \bibnamefont{Lenaghan}},
  \bibnamefont{and}
  \bibinfo{author}{\bibfnamefont{K.}~\bibnamefont{Splittorff}},
  \bibinfo{journal}{Phys. Rev. D} \textbf{\bibinfo{volume}{65}},
  \bibinfo{pages}{085015} (\bibinfo{year}{2002}).

\bibitem[{\citenamefont{Creutz}(2004)}]{Creutz:2003xu}
\bibinfo{author}{\bibfnamefont{M.}~\bibnamefont{Creutz}},
  \bibinfo{journal}{Phys. Rev. Lett.} \textbf{\bibinfo{volume}{92}},
  \bibinfo{pages}{201601} (\bibinfo{year}{2004}).

\bibitem[{\citenamefont{Metlitski and Zhitnitsky}(2005)}]{Metlitski:2005db}
\bibinfo{author}{\bibfnamefont{M.}~\bibnamefont{Metlitski}} \bibnamefont{and}
  \bibinfo{author}{\bibfnamefont{A.}~\bibnamefont{Zhitnitsky}},
  \bibinfo{journal}{Nucl. Phys. B} \textbf{\bibinfo{volume}{731}},
  \bibinfo{pages}{309} (\bibinfo{year}{2005}).

\bibitem[{\citenamefont{Metlitski and Zhitnitsky}(2006)}]{Metlitski:2005di}
\bibinfo{author}{\bibfnamefont{M.}~\bibnamefont{Metlitski}} \bibnamefont{and}
  \bibinfo{author}{\bibfnamefont{A.}~\bibnamefont{Zhitnitsky}},
  \bibinfo{journal}{Phys. Lett. B} \textbf{\bibinfo{volume}{633}},
  \bibinfo{pages}{721} (\bibinfo{year}{2006}).

\bibitem[{\citenamefont{Fujihara et~al.}(2007)\citenamefont{Fujihara, Inagaki,
  and Kimura}}]{fujihara:2005wk}
\bibinfo{author}{\bibfnamefont{T.}~\bibnamefont{Fujihara}},
  \bibinfo{author}{\bibfnamefont{T.}~\bibnamefont{Inagaki}}, \bibnamefont{and}
  \bibinfo{author}{\bibfnamefont{D.}~\bibnamefont{Kimura}},
  \bibinfo{journal}{Prog. Theor. Phys.} \textbf{\bibinfo{volume}{117}},
  \bibinfo{pages}{139} (\bibinfo{year}{2007}).

\bibitem[{\citenamefont{'t~Hooft}(1976)}]{tHooft:1976fv}
\bibinfo{author}{\bibfnamefont{G.}~\bibnamefont{'t~Hooft}},
  \bibinfo{journal}{Phys. Rev. D} \textbf{\bibinfo{volume}{14}},
  \bibinfo{pages}{3432} (\bibinfo{year}{1976}).

\bibitem[{\citenamefont{'t~Hooft}(1986)}]{tHooft:1986nc}
\bibinfo{author}{\bibfnamefont{G.}~\bibnamefont{'t~Hooft}},
  \bibinfo{journal}{Phys. Rep.} \textbf{\bibinfo{volume}{142}},
  \bibinfo{pages}{357} (\bibinfo{year}{1986}).

\bibitem[{\citenamefont{Cohen}(2001)}]{Cohen:2001hf}
\bibinfo{author}{\bibfnamefont{T.~D.} \bibnamefont{Cohen}},
  \bibinfo{journal}{Phys. Rev. D} \textbf{\bibinfo{volume}{64}},
  \bibinfo{pages}{047704} (\bibinfo{year}{2001}).

\bibitem[{\citenamefont{Einhorn and Wudka}(2003)}]{Einhorn:2002rm}
\bibinfo{author}{\bibfnamefont{M.~B.} \bibnamefont{Einhorn}} \bibnamefont{and}
  \bibinfo{author}{\bibfnamefont{J.}~\bibnamefont{Wudka}},
  \bibinfo{journal}{Phys. Rev. D} \textbf{\bibinfo{volume}{67}},
  \bibinfo{pages}{045004} (\bibinfo{year}{2003}).

\bibitem[{\citenamefont{Lee}(1973)}]{Lee:1973iz}
\bibinfo{author}{\bibfnamefont{T.~D.} \bibnamefont{Lee}},
  \bibinfo{journal}{Phys. Rev. D} \textbf{\bibinfo{volume}{8}},
  \bibinfo{pages}{1226} (\bibinfo{year}{1973}).

\bibitem[{\citenamefont{Morley and Schmidt}(1985)}]{Morley:1983wr}
\bibinfo{author}{\bibfnamefont{P.~D.} \bibnamefont{Morley}} \bibnamefont{and}
  \bibinfo{author}{\bibfnamefont{I.~A.} \bibnamefont{Schmidt}},
  \bibinfo{journal}{Z. Phys. C} \textbf{\bibinfo{volume}{26}},
  \bibinfo{pages}{627} (\bibinfo{year}{1985}).

\bibitem[{\citenamefont{Kharzeev et~al.}(1998)\citenamefont{Kharzeev, Pisarski,
  and Tytgat}}]{Kharzeev:1998kz}
\bibinfo{author}{\bibfnamefont{D.}~\bibnamefont{Kharzeev}},
  \bibinfo{author}{\bibfnamefont{R.~D.} \bibnamefont{Pisarski}},
  \bibnamefont{and} \bibinfo{author}{\bibfnamefont{M.~H.~G.}
  \bibnamefont{Tytgat}}, \bibinfo{journal}{Phys. Rev. Lett.}
  \textbf{\bibinfo{volume}{81}}, \bibinfo{pages}{512} (\bibinfo{year}{1998}).

\bibitem[{\citenamefont{Buckley et~al.}(2000)\citenamefont{Buckley, Fugleberg,
  and Zhitnitsky}}]{Buckley:1999mv}
\bibinfo{author}{\bibfnamefont{K.}~\bibnamefont{Buckley}},
  \bibinfo{author}{\bibfnamefont{T.}~\bibnamefont{Fugleberg}},
  \bibnamefont{and}
  \bibinfo{author}{\bibfnamefont{A.}~\bibnamefont{Zhitnitsky}},
  \bibinfo{journal}{Phys. Rev. Lett.} \textbf{\bibinfo{volume}{84}},
  \bibinfo{pages}{4814} (\bibinfo{year}{2000}).

\bibitem[{\citenamefont{Kharzeev and Pisarski}(2000)}]{Kharzeev:1999cz}
\bibinfo{author}{\bibfnamefont{D.}~\bibnamefont{Kharzeev}} \bibnamefont{and}
  \bibinfo{author}{\bibfnamefont{R.~D.} \bibnamefont{Pisarski}},
  \bibinfo{journal}{Phys. Rev. D} \textbf{\bibinfo{volume}{61}},
  \bibinfo{pages}{111901} (\bibinfo{year}{2000}).

\bibitem[{\citenamefont{Voloshin}(2004)}]{Voloshin:2004vk}
\bibinfo{author}{\bibfnamefont{S.~A.} \bibnamefont{Voloshin}},
  \bibinfo{journal}{Phys. Rev. C} \textbf{\bibinfo{volume}{70}},
  \bibinfo{pages}{057901} (\bibinfo{year}{2004}).

\bibitem[{\citenamefont{Kharzeev and Zhitnitsky}(2007)}]{Kharzeev:2007tn}
\bibinfo{author}{\bibfnamefont{D.}~\bibnamefont{Kharzeev}} \bibnamefont{and}
  \bibinfo{author}{\bibfnamefont{A.}~\bibnamefont{Zhitnitsky}},
  \bibinfo{journal}{Nucl. Phys. A} \textbf{\bibinfo{volume}{797}},
  \bibinfo{pages}{67} (\bibinfo{year}{2007}).

\bibitem[{\citenamefont{Kharzeev et~al.}(2008)\citenamefont{Kharzeev, McLerran,
  and Warringa}}]{Kharzeev:2007jp}
\bibinfo{author}{\bibfnamefont{D.~E.} \bibnamefont{Kharzeev}},
  \bibinfo{author}{\bibfnamefont{L.~D.} \bibnamefont{McLerran}},
  \bibnamefont{and} \bibinfo{author}{\bibfnamefont{H.~J.}
  \bibnamefont{Warringa}}, \bibinfo{journal}{Nucl. Phys. A}
  \textbf{\bibinfo{volume}{803}}, \bibinfo{pages}{227} (\bibinfo{year}{2008}).

\bibitem[{\citenamefont{Peccei and Quinn}(1977{\natexlab{a}})}]{Peccei:1977hh}
\bibinfo{author}{\bibfnamefont{R.~D.} \bibnamefont{Peccei}} \bibnamefont{and}
  \bibinfo{author}{\bibfnamefont{H.~R.} \bibnamefont{Quinn}},
  \bibinfo{journal}{Phys. Rev. Lett.} \textbf{\bibinfo{volume}{38}},
  \bibinfo{pages}{1440} (\bibinfo{year}{1977}{\natexlab{a}}).

\bibitem[{\citenamefont{Peccei and Quinn}(1977{\natexlab{b}})}]{Peccei:1977ur}
\bibinfo{author}{\bibfnamefont{R.~D.} \bibnamefont{Peccei}} \bibnamefont{and}
  \bibinfo{author}{\bibfnamefont{H.~R.} \bibnamefont{Quinn}},
  \bibinfo{journal}{Phys. Rev. D} \textbf{\bibinfo{volume}{16}},
  \bibinfo{pages}{1791} (\bibinfo{year}{1977}{\natexlab{b}}).

\bibitem[{\citenamefont{Wilczek}(1978)}]{Wilczek:1977pj}
\bibinfo{author}{\bibfnamefont{F.}~\bibnamefont{Wilczek}},
  \bibinfo{journal}{Phys. Rev. Lett.} \textbf{\bibinfo{volume}{40}},
  \bibinfo{pages}{279} (\bibinfo{year}{1978}).

\bibitem[{\citenamefont{Nambu and
  Jona-Lasinio}(1961{\natexlab{a}})}]{Nambu:1961tp}
\bibinfo{author}{\bibfnamefont{Y.}~\bibnamefont{Nambu}} \bibnamefont{and}
  \bibinfo{author}{\bibfnamefont{G.}~\bibnamefont{Jona-Lasinio}},
  \bibinfo{journal}{Phys Rev.} \textbf{\bibinfo{volume}{122}},
  \bibinfo{pages}{345} (\bibinfo{year}{1961}{\natexlab{a}}).

\bibitem[{\citenamefont{Nambu and
  Jona-Lasinio}(1961{\natexlab{b}})}]{Nambu:1961fr}
\bibinfo{author}{\bibfnamefont{Y.}~\bibnamefont{Nambu}} \bibnamefont{and}
  \bibinfo{author}{\bibfnamefont{G.}~\bibnamefont{Jona-Lasinio}},
  \bibinfo{journal}{Phys. Rev.} \textbf{\bibinfo{volume}{124}},
  \bibinfo{pages}{246} (\bibinfo{year}{1961}{\natexlab{b}}).

\bibitem[{\citenamefont{Buballa}(2005)}]{Buballa:2003qv}
\bibinfo{author}{\bibfnamefont{M.}~\bibnamefont{Buballa}},
  \bibinfo{journal}{Phys. Rept.} \textbf{\bibinfo{volume}{407}},
  \bibinfo{pages}{205} (\bibinfo{year}{2005}).

\bibitem[{\citenamefont{Fujikawa}(1979)}]{Fujikawa:1979ay}
\bibinfo{author}{\bibfnamefont{K.}~\bibnamefont{Fujikawa}},
  \bibinfo{journal}{Phys. Rev. Lett.} \textbf{\bibinfo{volume}{42}},
  \bibinfo{pages}{1195} (\bibinfo{year}{1979}).

\bibitem[{\citenamefont{Frank et~al.}(2003)\citenamefont{Frank, Buballa, and
  Oertel}}]{Frank:2003ve}
\bibinfo{author}{\bibfnamefont{M.}~\bibnamefont{Frank}},
  \bibinfo{author}{\bibfnamefont{M.}~\bibnamefont{Buballa}}, \bibnamefont{and}
  \bibinfo{author}{\bibfnamefont{M.}~\bibnamefont{Oertel}},
  \bibinfo{journal}{Phys. Lett. B} \textbf{\bibinfo{volume}{562}},
  \bibinfo{pages}{221} (\bibinfo{year}{2003}).

\bibitem[{\citenamefont{Warringa et~al.}(2005)\citenamefont{Warringa, Boer, and
  Andersen}}]{Warringa:2005jh}
\bibinfo{author}{\bibfnamefont{H.~J.} \bibnamefont{Warringa}},
  \bibinfo{author}{\bibfnamefont{D.}~\bibnamefont{Boer}}, \bibnamefont{and}
  \bibinfo{author}{\bibfnamefont{J.~O.} \bibnamefont{Andersen}},
  \bibinfo{journal}{Phys. Rev. D} \textbf{\bibinfo{volume}{72}},
  \bibinfo{pages}{014015} (\bibinfo{year}{2005}).

\bibitem[{\citenamefont{Warringa}(2006)}]{Warringa:2005my}
\bibinfo{author}{\bibfnamefont{H.~J.} \bibnamefont{Warringa}}, \eprint{hep-ph/0606063}.

\bibitem[{\citenamefont{Gross et~al.}(1981)\citenamefont{Gross, Pisarski, and
  Yaffe}}]{Gross:1980br}
\bibinfo{author}{\bibfnamefont{D.~J.} \bibnamefont{Gross}},
  \bibinfo{author}{\bibfnamefont{R.~D.} \bibnamefont{Pisarski}},
  \bibnamefont{and} \bibinfo{author}{\bibfnamefont{L.~G.} \bibnamefont{Yaffe}},
  \bibinfo{journal}{Rev. Mod. Phys.} \textbf{\bibinfo{volume}{53}},
  \bibinfo{pages}{43} (\bibinfo{year}{1981}).

\bibitem[{\citenamefont{Son and Stephanov}(2001)}]{Son:2000xc}
\bibinfo{author}{\bibfnamefont{D.~T.} \bibnamefont{Son}} \bibnamefont{and}
  \bibinfo{author}{\bibfnamefont{M.~A.} \bibnamefont{Stephanov}},
  \bibinfo{journal}{Phys. Rev. Lett.} \textbf{\bibinfo{volume}{86}},
  \bibinfo{pages}{592} (\bibinfo{year}{2001}).

\bibitem[{\citenamefont{Barducci et~al.}(2004)\citenamefont{Barducci,
  Casalbuoni, Pettini, and Ravagli}}]{Barducci:2004tt}
\bibinfo{author}{\bibfnamefont{A.}~\bibnamefont{Barducci}},
  \bibinfo{author}{\bibfnamefont{R.}~\bibnamefont{Casalbuoni}},
  \bibinfo{author}{\bibfnamefont{G.}~\bibnamefont{Pettini}}, \bibnamefont{and}
  \bibinfo{author}{\bibfnamefont{L.}~\bibnamefont{Ravagli}},
  \bibinfo{journal}{Phys. Rev. D} \textbf{\bibinfo{volume}{69}},
  \bibinfo{pages}{096004} (\bibinfo{year}{2004}).

\bibitem[{\citenamefont{Barducci et~al.}(2005)\citenamefont{Barducci,
  Casalbuoni, Pettini, and Ravagli}}]{Barducci:2004nc}
\bibinfo{author}{\bibfnamefont{A.}~\bibnamefont{Barducci}},
  \bibinfo{author}{\bibfnamefont{R.}~\bibnamefont{Casalbuoni}},
  \bibinfo{author}{\bibfnamefont{G.}~\bibnamefont{Pettini}}, \bibnamefont{and}
  \bibinfo{author}{\bibfnamefont{L.}~\bibnamefont{Ravagli}},
  \bibinfo{journal}{Phys. Rev. D} \textbf{\bibinfo{volume}{71}},
  \bibinfo{pages}{016011} (\bibinfo{year}{2005}).

\bibitem[{\citenamefont{He and Zhuang}(2005)}]{He:2005sp}
\bibinfo{author}{\bibfnamefont{L.}~\bibnamefont{He}} \bibnamefont{and}
  \bibinfo{author}{\bibfnamefont{P.}~\bibnamefont{Zhuang}},
  \bibinfo{journal}{Phys. Lett. B} \textbf{\bibinfo{volume}{615}},
  \bibinfo{pages}{93} (\bibinfo{year}{2005}).

\bibitem[{\citenamefont{Klevansky}(1992)}]{Klevansky:1992qe}
\bibinfo{author}{\bibfnamefont{S.}~\bibnamefont{Klevansky}},
  \bibinfo{journal}{Rev. Mod. Phys.} \textbf{\bibinfo{volume}{64}},
  \bibinfo{pages}{649} (\bibinfo{year}{1992}).

\end{thebibliography}

\end{document}